\documentclass[12pt]{iopart}
\usepackage{epsfig}

\newenvironment{color}[3]
{
}{
}
\newcommand {\red}[1]       {#1}

\newcommand {\grey}[1]      {}

\newcommand{\pT}{p_T}
\newcommand{\pTt}{p_T^{(t)}}
\newcommand{\pTa}{p_T^{(a)}}
\newcommand{\dphi}{\Delta\phi}
\newcommand{\deta}{\Delta\eta}
\newcommand {\phis}	{\phi_{s}}

\newcommand {\psiEP}	{\psi_{EP}}

\newcommand {\vf}	{v_2}
\newcommand {\vv}	{v_4}

\newcommand {\vAS}	{$v_2$\{2,{\sc as}\}}
\newcommand {\flow}[1]	{v_2\{#1\}}
\newcommand {\mean}[1]	{\langle #1\rangle}

\newcommand {\zyam}	{{\sc zyam}}
\newcommand {\gev}	{{GeV/$c$}}

\begin{document}

\title[Geometrical Effect on Conical Emission of Correlated Hadrons]{Geometrical Effect on Conical Emission of Correlated Hadrons}

\author{Fuqiang Wang (for the STAR Collaboration)}

\address{Department of Physics, Purdue University, West Lafayette, Indiana 47907, USA}
\ead{fqwang@purdue.edu}
\begin{abstract}
Dihadron correlations at intermediate $\pT$ revealed novel structures on the away side of high $\pT$ trigger particles at RHIC. The away-side correlations in central Au+Au collisions are significantly broader than in pp and d+Au collisions and in restricted kinematic range, double-peaked away from $\dphi=\pi$. Three-particle correlations indicate conical emission of the away-side correlated hadrons at angles independent of associated particle $\pT$, consistent with formation of Mach-cone shock waves. In this talk we further investigate the conical emission phenomenon exploiting dihadron correlations as a function of the trigger particle azimuth from the reaction plane. Such correlations are sensitive to the collision geometry.
% thereby providing unique information on away-side conical emission. 
We study these geometrical effects and discuss how they might be used to further our understanding of the medium created in heavy-ion collisions.%, such as its speed of sound and equation of state.
\end{abstract}

%Uncomment for PACS numbers title message
%\pacs{00.00, 20.00, 42.10}
% Keywords required only for MST, PB, PMB, PM, JOA, JOB? 
%\vspace{2pc}
%\noindent{\it Keywords}: Article preparation, IOP journals
% Uncomment for Submitted to journal title message
%\submitto{\JPA}
% Comment out if separate title page not required
%\maketitle

\section{Introduction}

Dihadron jet-like correlations with a high transverse momentum ($\pT$) trigger particle provide a unique tool to study the hot and dense medium created in relativistic heavy-ion collisions, owing to the fact that the away-side partner jet has to traverse the entire medium due to the surface bias of the production points of high-$\pT$ particles. Dihadron correlations of intermediate $\pT$ associated particles revealed novel structures on the away side of the high $\pT$ trigger particle. The away-side correlations in central Au+Au collisions are significantly broader than in pp and d+Au collisions and in restricted kinematic range, double-peaked away from $\dphi=\pi$~\cite{jetspec,Horner,Phenix}. This observation has motivated many theoretical investigations of its physics origin~\cite{Stoecker,Casalderrey,Ruppert,Renk}.
Three-particle correlations showed evidence of conical emission of away-side correlated hadrons~\cite{3part}. The conical emission angle is found to be independent of the associated particle $\pT$, suggesting the underlying physics mechanism may be Mach-cone shock waves. Three-particle cumulant in azimuth relative to the reaction plane~\cite{3part} confirms this finding. Recent study~\cite{Takahashi} suggests that fluctuations in initial conditions may also create an away-side double-hump structure in dihadron correlations. Such mechanisms, however, would not generate conical emission structures in three-particle correlations~\cite{Qian}.

%\begin{figure}
%\begin{center}
%\includegraphics[width=0.90\textwidth]{cumulant.eps}
%%\framebox(200,150){cumulant plots}
%\end{center}
%\caption{Three-particle cumulant measurements.}
%\label{fig2}
%\end{figure}

The conical emission angle is measured to be 
$\theta = 1.37 \pm 0.02 {\rm (stat.)} \pm 0.06 {\rm (syst.)}$~\cite{3part}. 
If Mach cones are indeed the underlying mechanism, one may obtain, ideally, the medium's speed of sound via $c_s=\cos(\theta)$. However, model studies~\cite{Renk} indicate that the Mach cone angle would be strongly affected by the medium expansion. The effects depend on the relative configurations of the Mach cones and the collision geometry. We attempt to study these effects by exploiting dihadron correlations as a function of the trigger particle azimuth relative to the reaction plane (RP). Such correlations are sensitive to the collision geometry as well as the orientation of the possible Mach cones. 
%We study these geometrical effects and discuss how they might be used to further our understanding of the medium created in heavy-ion collisions, such as its speed of sound and equation of state.

\section{Dihadron correlation analysis relative to the reaction plane}

We use the second Fourier harmonic to determine the event plane (EP) azimuth $\psiEP$~\cite{flowMethod}, using particles below $\pT = 2$~\gev. To avoid self-correlation, particles from the $\pT$ bin used for our correlation analysis are excluded from the EP construction. Non-flow correlations, such as dijets, can influence the EP determination. To reduce this effect, we use the modified reaction plane ({\sc mrp}) method~\cite{v2MRP}, excluding particles within $|\deta|<0.5$ of the highest $\pT$ particle in the event from the EP construction.
%We use the $\pT$-weight method~\cite{flowMethod} which gives better EP resolution due to the stronger anisotropy at larger $\pT$. The slight non-uniform efficiency and acceptance in azimuthal angle were corrected for in the event-plane construction. The constructed EP $\psiEP$ is uniformly distributed.

We divide our data into six slices in $\phis$, the trigger particle azimuth relative to the EP, and analyze azimuthal correlations separately in each slice. 
%The correlation sits atop a large background. 
%The background has a flow modulation given~\cite{Bielcikova} by
%%\begin{equation}
%%\frac{dN}{d\dphi}=B\left[1+2\vf\vf^R\cos(2\dphi)+2\vv\vv^R\cos(4\dphi)\right],
%%\label{eq:bkgd}
%%\end{equation}
%$dN/d\dphi=B\left[1+2\vf\vf^R\cos(2\dphi)+2\vv\vv^R\cos(4\dphi)\right]$,
%where $\vf$ and $\vv$ are the associated particle's second and fourth harmonics, and $\vf^R$ and $\vv^R$ are those of the trigger particles, 
%$\vf^R=\left\langle\cos\left[2\left(\phit-\psiRP\right)\right]\right\rangle^R$ and
%$\vv^R=\left\langle\cos\left[4\left(\phit-\psiRP\right)\right]\right\rangle^R$,
%averaged within a slice of width $2c$ at $\phis$, 
%$\phis-c<\left|\phit-\psiRP\right|<\phis+c$ ($c=\pi/24$ in our analysis).
The correlation background has a flow modulation that depends on the trigger and associated particle's second and fourth harmonic anisotropies, $\vf$ and $\vv$, and the EP resolutions~\cite{Bielcikova}.
We obtain the EP resolutions by the sub-event method~\cite{flowMethod}. 
There are several measurements of elliptic flow anisotropies. %; many of them are affected, to various degrees, by non-flow contributions that are caused by particle correlations unrelated to the RP, such as resonance decays and (mini)jet-correlations. 
The two-particle and the {\sc mrp} methods give similar results and significantly overestimate elliptic flow due to non-flow~\cite{v2MRP}. The major component of non-flow is the measured small-angle minijet correlation~\cite{minijet}. 
%Since the elliptic flow effect is symmetric between near- and away-side, \vAS $=\sqrt{\mean{\cos2\dphi}}$ from only away-side pairs contains much less non-flow. Therefore we use \vAS, computed from untriggered two-particle azimuthal correlations from inclusive events for our centrality and $\pT$ bins, as our upper limit of systematic uncertainty. 
Since away-side pairs contain much less non-flow and the elliptic flow effect is symmetric between near- and away-side, we use \vAS $=\sqrt{\mean{\cos2\dphi}}$, computed from untriggered two-particle azimuthal correlations in inclusive events for our centrality and $\pT$ bins, as our upper limit of systematic uncertainty. 
The four particle method ($\flow{4}$)~\cite{Aihong} gives the smallest anisotropy parameter for the centrality range 20-60\% used in the present study~\cite{Kettler}. The $\flow{4}$ is likely an underestimate of elliptic flow because the flow fluctuation effect is negative in $\flow{4}$. We note that $\flow{4}$ may still contain some non-flow effects, however the agreement between $\flow{4}$ and the Lee-Yang-Zero method suggests that such non-flow effects are small. 
%$\vf$\{{\sc 2d}\} should in principle be a better estimate of the elliptic flow to be subtracted from the jet-correlation~\cite{Wang_flowbg}, however, decomposition of low $\pT$ two-particle correlation using a fit function sometimes leads to large uncertainties on the extracted quadrupole component because of the entangled fit parameters. 
%
We take $\vf=($\vAS$+\flow{4})/2$ and the range between \vAS\ and $\flow{4}$ as our uncertainty.
%We parameterized the $\vv$ measurement~\cite{v2MRP} within the measured $\pT$ range of 1-3~\gev\  as $\vv=1.15\vf^2$ and used this parameterization for both trigger and associated particles in our flow correction. 
We parameterized the $\vv$ measurement~\cite{v2MRP} by $\vv=1.15\vf^2$.

The background level $B$ is normalized using the Zero-Yield-At-Minimum (\zyam) method. The background levels can be different for the different $\phis$ slices because of the net effect of the variations in jet-quenching with $\phis$ and the centrality cuts on particle multiplicity in $|\eta|<0.5$. In our correlation analysis $B$ is treated independently in individual $\phis$ slices. The systematic uncertainty of $B$ due to \zyam\ itself is assessed by varying the size of the $\dphi$ normalization range between $\pi/12$ and $\pi/6$. The systematic uncertainty due to deviation of $B$ from \zyam\ is assessed by Gaussian fits to the \zyam-subtracted correlation functions, similar to those in Fig.~\ref{fig:corr_symm} but with a free pedestal, as well as by comparing $B$ to those obtained from asymmetric correlation functions (see Fig.~\ref{fig:corr_asym}). The effect of the $B$ uncertainty is an approximate constant shift to the baseline of the correlation functions, without significant change to their shapes. %We therefore do not dipict the \zyam\ background uncertainties in this proceedings.
The systematic uncertainty on $B$ is not included in the results reported in these proceedings.

%The \zyam\ assumption gives an upper limit to the underlying background level. One could make a better assessment with more stringent requirements, such as using three-particle correlation \zyam~\cite{3part}. Similarly, the background levels obtained by separately analyzing positive $\phis$ and negative $\phis$ are smaller than our default $B$ from \zyam\ of the combined correlation function of positive and negative $\phis$'s. We assign the difference in the obtained $B$ as additional, one-sided systematic uncertainty. 

When assessing the systematic uncertainty due to elliptic flow on our correlation results, we use \zyam\ to adjust background normalization. Due to the interplay between flow modulation and \zyam\ normalization, the uncertainty on our correlation functions due to flow is larger for in-plane than out-of-plane trigger particles (see Figs.~\ref{fig:corr_symm} and~\ref{fig:corr_asym}).
%Overall, the $\vf$ uncertainty due to non-flow has the largest effect on the systematic uncertainties of the final correlation results.

\red{Further details of the analysis can be found in Ref.~\cite{corrRP}.}

\section{Dihadron correlation results relative to the reaction plane}

\begin{figure}[hbt]
\begin{center}
\hfill
\includegraphics[width=0.98\textwidth]{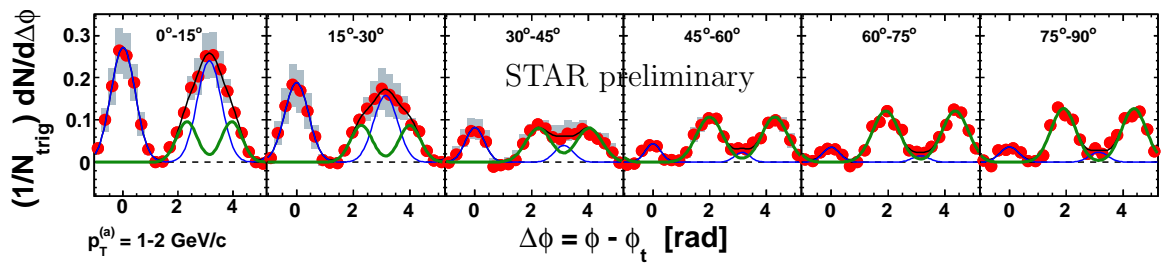}
\end{center}
\vspace{-3.3cm}\hfill STAR preliminary\hspace*{5.3cm}
\vspace{2.1cm}
\caption{Azimuthal correlation at $|\deta|>0.7$ versus trigger azimuth relative to RP, $|\phis|$, in 20-60\% Au+Au collisions\red{~\cite{corrRP}}. The trigger and associated $\pT$ ranges are $3<\pTt<4$~\gev\ and $1<\pTa<2$~\gev. The shaded areas are systematic uncertainties due to flow. The curves are Gaussian fit result: back-to-back ridges at $\dphi=0$ and $\pi$, and away-side peaks symmetric about $\dphi=\pi$.}
\label{fig:corr_symm}
\end{figure}

In this work we focus on the away-side dihadron correlations. We avoid the near-side jet-like component by analyzing azimuthal correlations at large $|\deta|>0.7$. Figure~\ref{fig:corr_symm} shows the azimuthal correlations in six slices of $|\phis|$ from in- to out-of-plane\red{~\cite{corrRP}}. The positive and negative $\phis$ slices are combined. The near-side peak is mainly due to the ridge which decreases from in-plane to out-of-plane. %\grey{~\cite{Aoqi}}. 
The away-side structure changes dramatically, from singly peaked in-plane to distinctively double-peaked out-of-plane. It appears that whenever there is a large ridge on the near side, there is a comparable component at $\dphi=\pi$ on the away side.

Many experimental observations suggest that the ridge and the jet-like component are unrelated~\cite{Puschke,Netrakanti,Nattrass}.
%On the other hand, there is much evidence to suggest that the ridge and the jet are unrelated and the conjectured correlation between the ridge and a high-$\pT$ trigger particle is accidental. For example, particle composition in the ridge is similar to that of the bulk medium. 
If they are indeed unrelated, then due to symmetry there ought to be an identical excess of momentum on the away-side to balance the ridge. This momentum balance is different from statistical momentum balance to an extra, fluctuating high-$\pT$ particle.  In other words, there is no criterion to determine which is near-side and which is away-side if the ridge and the jet are unrelated; the two sides have to be considered equally. 
It is thus attempting to fit 
%Thus the large $\deta$ azimuthal correlation functions (in the upper panel of Fig.~\ref{fig:jetRidge}) should be fit with two identical ridge Gaussian peaks (one at $\dphi=0$ and the other at $\dphi=\pi$) and two symmetric Gaussian peaks on the away side about $\dphi=\pi$. However, close inspection of the correlation in Fig.~\ref{fig:jetRidge} indicates that away-side $\pi$-region magnitude is smaller than the near-side ridge magnitude. The relative strength of the away-side to near-side ridge may depend on the associated $\pT$. This may be the reason why the difference between away-side $\pi$-region and the near-side ridge is even softer than the inclusive hadron spectrum. Indeed, our four-Gaussian fit with two identical back-to-back ridges and two symmetric cone peaks gives a relatively large $\chi^2/${\sc ndf}. Thus, we fit 
the large $\deta$ azimuthal correlation with two away-side Gaussians symmetric about $\dphi=\pi$ and two ridge Gaussians back-to-back at $\dphi=0$ and $\pi$. We keep identical width for the ridge Gaussians but allow their magnitudes to vary independently. The fit results are superimposed in Fig.~\ref{fig:corr_symm}. 
%The ratio of the away-side ridge magnitude to that of the near-side ridge is shown in Fig.~\ref{fig:rel_ridge} as a function of $\phis$ for associated $1<\pTa<2$~\gev\ and as a function of $\pTa$ for integrated $\phis$. The trigger particle $\pT$ is $3<\pTt<4$~\gev. The away- to near-side ridge ratio is generally smaller than unity, and decreases somewhat from in-plane to out-of-plane and decreases with associated $\pT$.

\begin{figure}[hbt]
\begin{center}
\hfill
\includegraphics[width=0.45\textwidth]{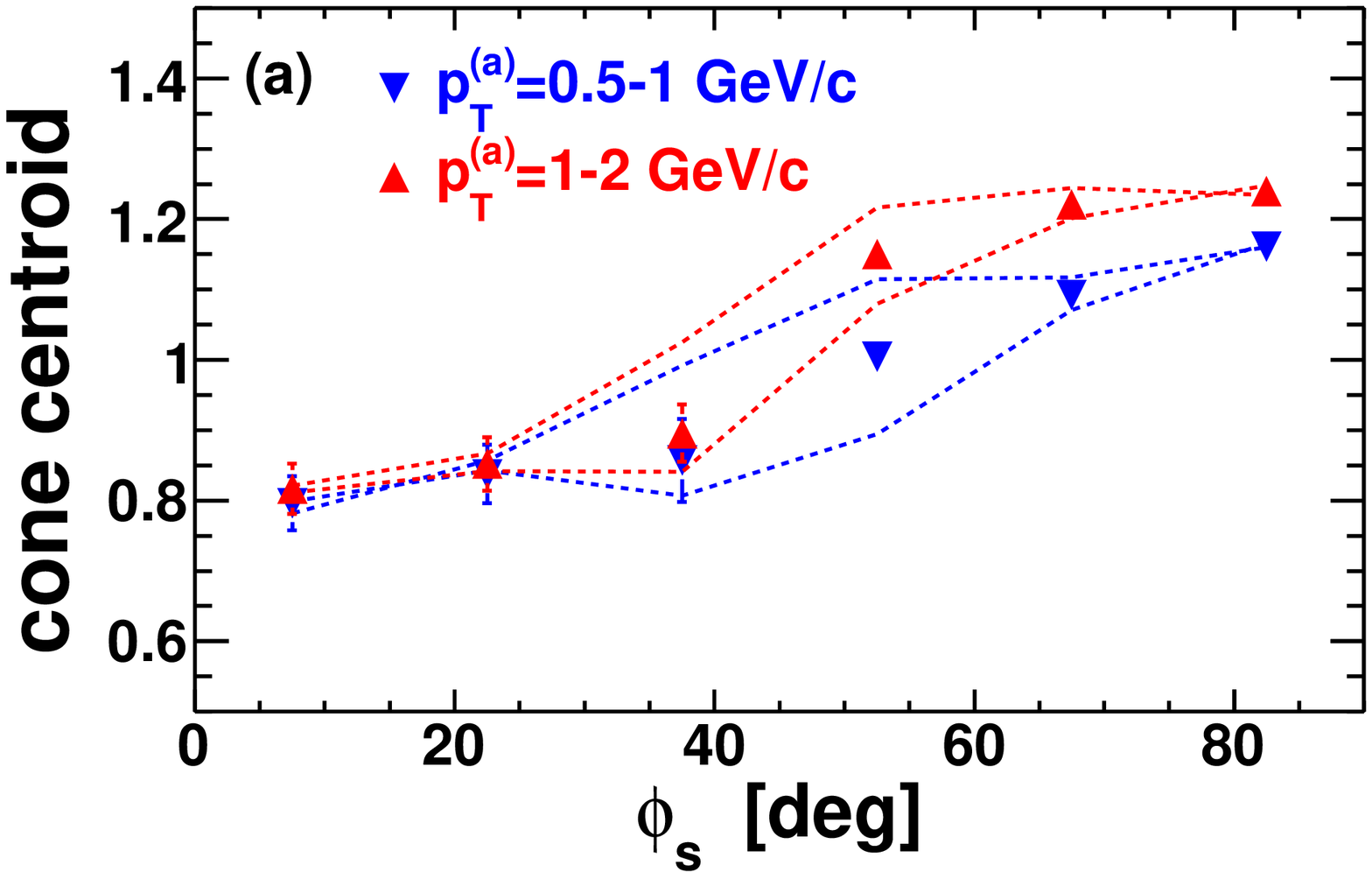}
\includegraphics[width=0.45\textwidth]{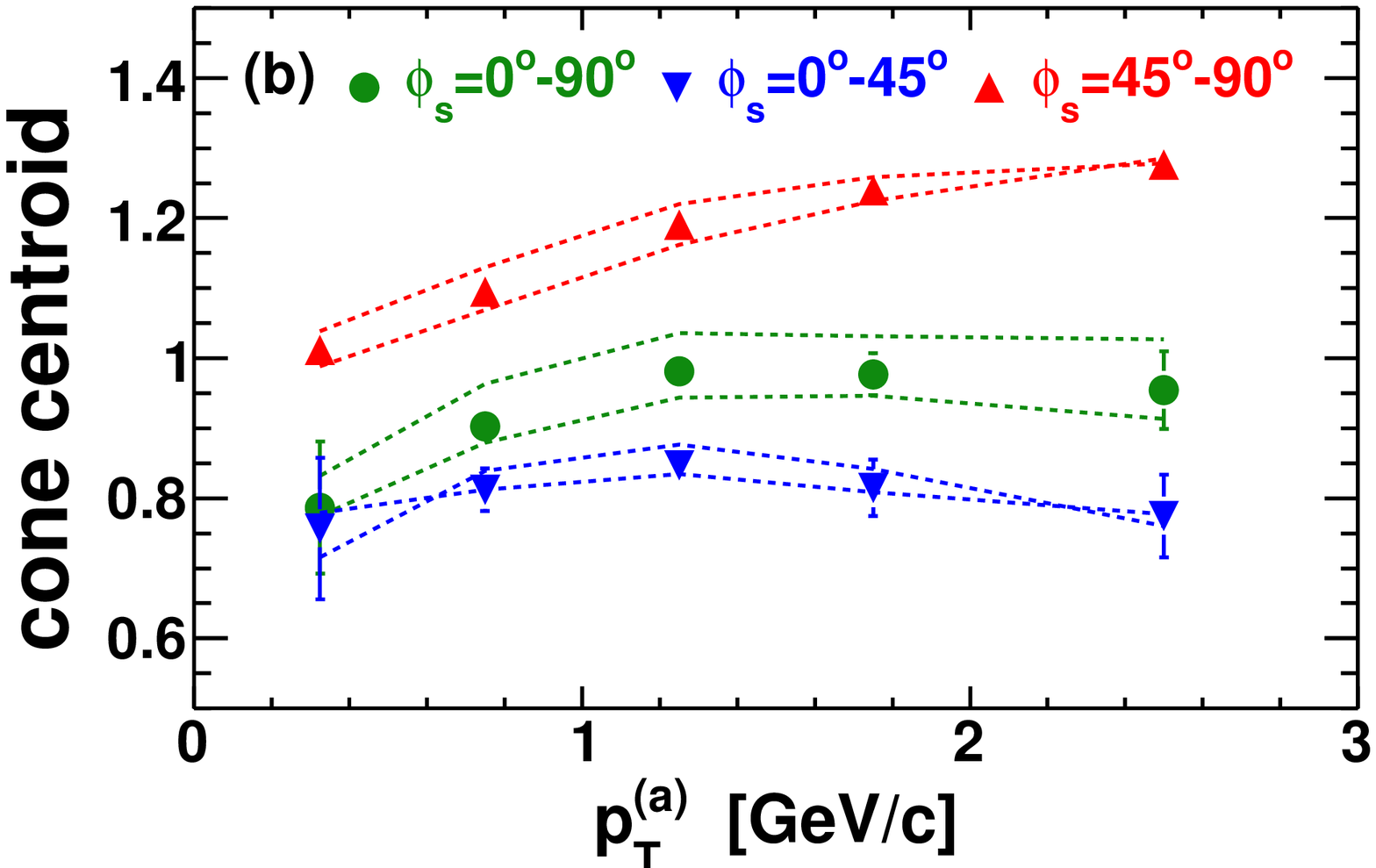}
\end{center}
%\begin{center}
%\vspace{-4.7cm}\hfill (a)\hspace{6.2cm}(b)\hspace*{3cm}
%\end{center}
%\vspace{1.8cm}\hfill STAR preliminary\hspace{3.9cm}STAR preliminary\hspace*{0.5cm}
\vspace{-1.9cm}\hfill STAR preliminary\hspace{3.9cm}STAR preliminary\hspace*{0.5cm}
\vspace{0.8cm}
\caption{Away-side double-peak fit position (relative to $\pi$) as a function of (a) $\phis$ and (b) $\pTa$. Error bars are statistical only. \red{The systematic uncertainties due to elliptic flow are indicated by the dashed lines.} Data are from 20-60\% Au+Au collisions, and the trigger particle $\pTt$ range is $3<\pTt<4$~\gev\red{~\cite{corrRP}}.}
\label{fig:corr_symm_fit}
\end{figure}

%We study the conical emission peak position obtained from our 4-Gaussian fit. 
Figure~\ref{fig:corr_symm_fit}(a) shows the fit double-peak angle as a function of $\phis$ for three associated $\pTa$ bins\red{~\cite{corrRP}}. The peak angle is approximately constant for $|\phis|<45^{\circ}$. For $|\phis|>45^{\circ}$ it increases with $\phis$, and becomes different for low and high $\pTa$. The larger angle for out-of-plane trigger particles may be due to a more significant influence from medium flow. For in-plane orientation, the conical emission hadrons on the away side are likely aligned with the medium flow direction, receiving insignificant deflection to their $\pT$. The conical emission angle may even be shrunk if it forms after passing the center of the medium. Moreover, the overlap collision zone is thinner in-plane, so the away-side correlated hadrons can escape the collision zone more easily. For out-of-plane orientation, on the other hand, the conical emission hadrons move more or less perpendicularly to the medium flow direction because of the long path length they have to traverse. They receive a large side-kick from the medium flow, broadening their final emission angle. 

Figure~\ref{fig:corr_symm_fit}(b) shows the conical emission peak angle as a function of $\pTa$ for in- and out-of-plane trigger particle orientations\red{~\cite{corrRP}}. The peak angle is relatively independent of associated $\pTa$ for in-plane trigger particles. %They may reflect more closely the average emission angle of correlated away-side hadrons because the medium flow effect is expected to be small as discussed above. 
The peak angle for the out-of-plane orientation is larger, consistent with a larger broadening effect from medium flow. However, the angle position increases with $\pTa$, which is naively not expected if those particles are pushed by the same medium flow velocity. It is, however, possible that the higher $\pTa$ hadrons are emitted earlier by the away-side parton while traversing the medium, thereby receiving a larger flow effect in the outer region of the bulk medium than the low $\pTa$ hadrons~\cite{Ma}. 

It is worthwhile to note that the peak positions reported here are from fits to dihadron correlations. They are different from those obtained from three-particle correlations~\cite{3part} where the conical emission angle was found to be independent of associated particle $\pTa$. The angle from the three-particle correlation fit is cleaner because the peaks are more cleanly separated in the two-dimensional azimuth space, while the fit to dihadron correlations is more affected by other physics effects. One such effect is jet deflection, which was found to be present in three-particle correlation where the diagonal peaks are stronger than the off-diagonal conical emission peaks~\cite{3part}.

\begin{figure}[hbt]
\begin{center}
\hfill
\includegraphics[width=0.98\textwidth]{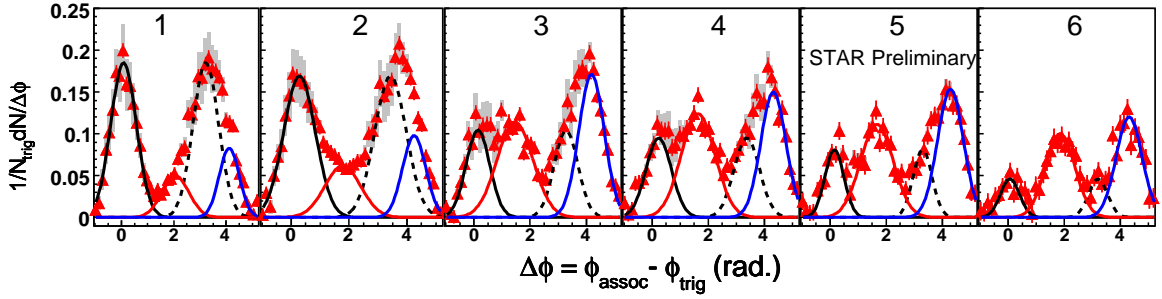}
\end{center}
\vspace{-0.7cm}
\caption{Same as Fig.~\ref{fig:corr_symm}, except that the correlation in a positive $\phis$ bin is flipped and properly shifted before combined with that in the symmetric negative $\phis$ bin. The away-side peaks are now not symmetric about $\dphi=\pi$. The associated $\pTa$ range is $1<\pTa<1.5$~\gev.}
\label{fig:corr_asym}
\end{figure}

The results reported above combine the trigger particles above and below the EP. We can study the correlations of those trigger particles separately and learn more about the interplay between jet-correlation and the collision geometry. Those results are shown in Fig.~\ref{fig:corr_asym}, where the correlation for a positive $\phis$ bin is flipped via $\dphi\rightarrow-\dphi$ and properly shifted before combined with that in the symmetric negative $\phis$ bin. The background normalization is done by the \zyam\ method; the subtracted background is lower than in Fig.~\ref{fig:corr_symm} because the minimum now appears at only one side, $\dphi\approx-1$ instead of both sides ($\dphi\approx\pm1$) in Fig.~\ref{fig:corr_symm}. This difference is used as an assessment of the \zyam\ uncertainty as aforementioned. Again we fit the correlation functions with four Gaussians: two for the back-to-back ridges and two for the away-side conical emission peaks. We fix the back-to-back ridges to be identical including their amplitudes. We also allowed their magnitudes to vary independently, and obtained consistent fit results.

\begin{figure}[hbt]
\vspace*{-0.5cm}
\begin{center}
\hfill
\includegraphics[width=0.73\textwidth]{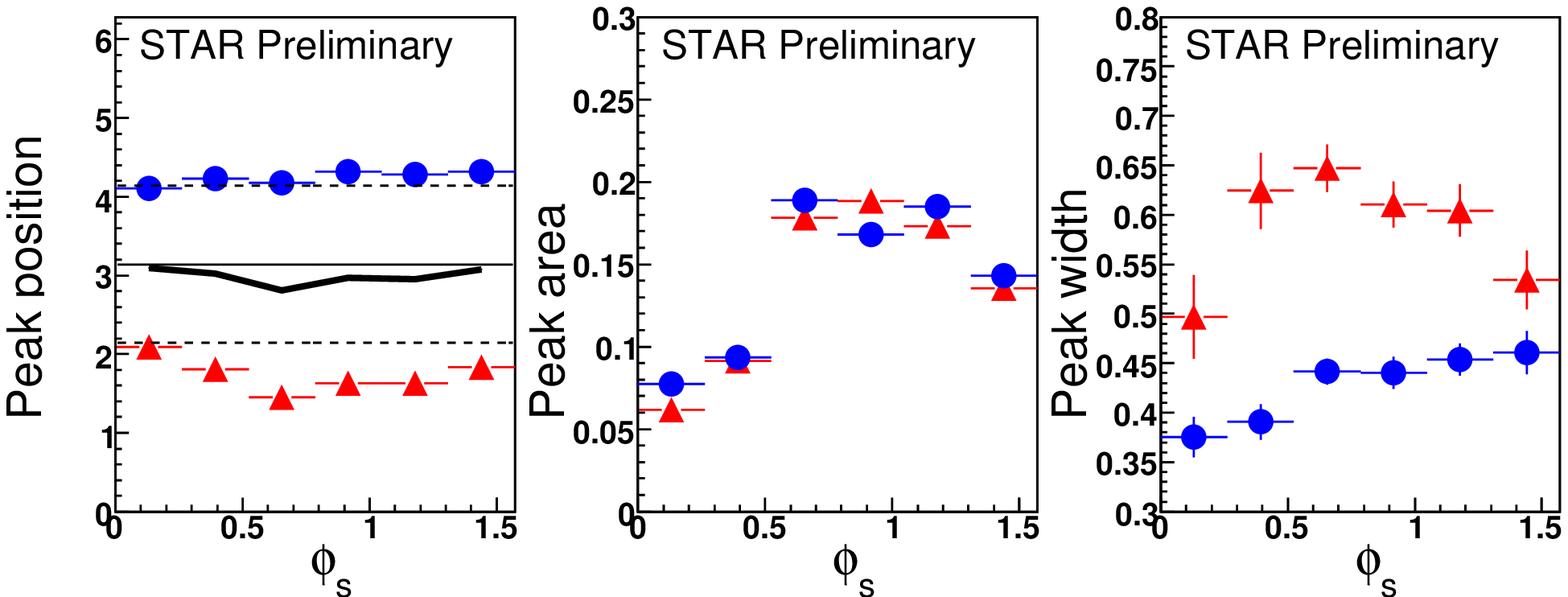}
\includegraphics[width=0.25\textwidth]{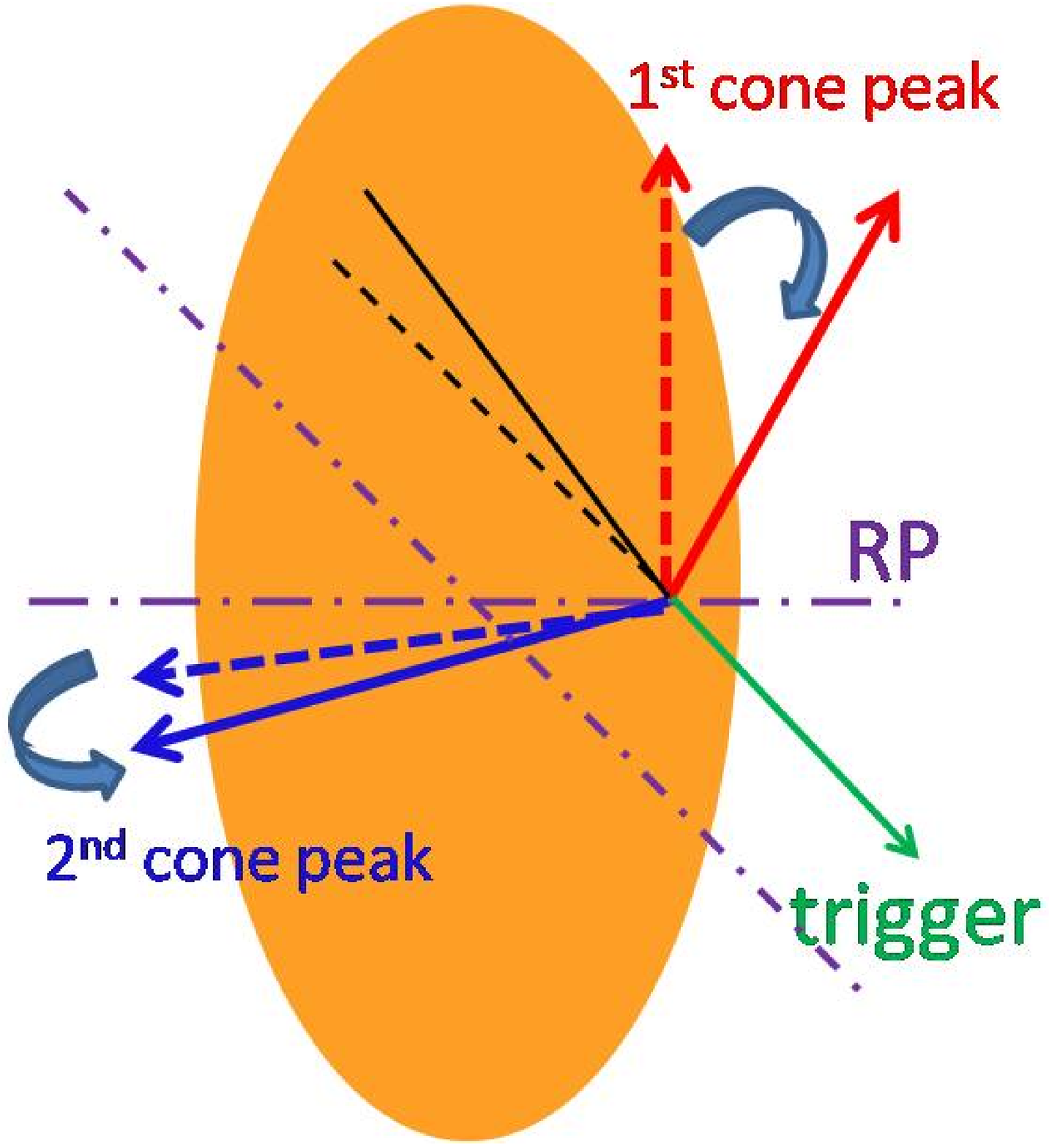}
\end{center}
%\vspace{-3.7cm}\hfill (a)\hspace{2.5cm}(b)\hspace{2.5cm}(c)\hspace{2.5cm}(d)\hspace*{2cm}
\vspace{-0.7cm}
\caption{The away-side double-peak (a) positions, (b) areas, and (c) widths from the fits in Fig.~\ref{fig:corr_asym}. (d) Illustration of away-side conical emission angles. The triangles (circles) in (a-c) correspond to the first (second) conical emission peak in (d).}
\label{fig:corr_asym_fit}
\end{figure}

Figure~\ref{fig:corr_asym_fit}(a) shows the away-side peak positions versus $\phis$. The cartoon aids to the visualization. As discussed above, the conical emission angle is not significantly affected by medium flow when the trigger particle is aligned in the RP. As the trigger particle moves away from the RP, the angle of the second peak (circles) does not seem to change, indicating insignificant effect from medium flow. In contrary, the other conical emission peak changes its location significantly. The largest change appears at $\phis\sim45^\circ$.
%These data may indicate evidence of flow effect on conical emission. 
If the away-side partner parton is deflected by the medium flow and then lose energy generating conical emission at a fixed angle, then the relative distance between the peaks would be the same. The data seem to indicate otherwise, suggesting that it is the conical emission particles that are pushed by flow, not the away-side partner parton.

It was suggested that asymmetric away-side peaks may arise from absorption of particles traversing different amount of medium~\cite{Jia}. To investigate such effect, we show in Fig.~\ref{fig:corr_asym_fit}(b,c) the Gaussian areas and widths of the away-side peaks. The peak areas are similar, without evidence of medium absorption, whereas the peak widths are broadened, consistent with broadening due to the medium flow.

\section{Summary}
We have studied dihadron correlations with high $\pT$ trigger particles at $|\deta|>0.7$ as a function of the trigger azimuth ($\phis$) relative to the event plane. We combine the correlations from the two symmetric $\phis>0$ and $\phis<0$ bins straightforwardly as well as after flipping the former in $\dphi$. We argue the near-side ridge is accompanied by a similar ridge on the away side. We fit the correlation function with four Gaussians, two for the back-to-back ridges and two for the away-side conical emission peaks. We investigate the fit parameters as a function of $\phis$. We found significant variations in the peak positions, areas, widths with $\phis$, suggesting geometrical effects on conical emission. These effects are likely due to the medium flow. We found no evidence of medium absorption. Our study should help disentangle medium flow effects in the conical emission signal and may further our understanding of the medium created in heavy-ion collisions, such as its speed of sound and equation of state. \grey{One of the future tasks is to assess the systematic effects of elliptic flow and \zyam\ background uncertainties on our fit results.}

%\section*{Acknowledgment}

\end{document}